\documentclass[12pt,a4paper]{revtex4-2}
\usepackage[utf8x]{inputenc}
\usepackage{ucs}
\usepackage[english]{babel}
\usepackage{amsmath}
\usepackage{amsfonts}
\usepackage{amssymb}
\usepackage{graphicx}
\usepackage[left=2cm,right=2cm,top=2cm,bottom=2cm]{geometry}

\usepackage{bm}

\usepackage{indentfirst}

\usepackage{float}
\usepackage[caption = false]{subfig}
\begin{document}
\title{Wigner current in multidimensional quantum billiards}

\author{S.S. Seidov}
\affiliation{HSE University, Moscow, Russia, alikseidov@yandex.ru}

\author{D.G. Bezymiannykh}
\affiliation{HSE University, Moscow, Russia}

\newcommand{\vx}{\bm{x}}
\newcommand{\vp}{\bm{p}}
\newcommand{\vy}{\bm{y}}
\newcommand{\vj}{\bm{j}}
\newcommand{\vk}{\bm{k}}

\renewcommand{\Im}{\operatorname{Im}}
\renewcommand{\Re}{\operatorname{Re}}

\newcommand{\F}{\mathcal{F}}

\begin{abstract}
In the present paper we derive the Wigner current of the particle in a multidimensional billiard --- the compact region of space in which the particle moves freely. The calculation is based on proposed by us previously method of imposing boundary conditions by convolution of the free particle Wigner function with some time independent function, defined by the shape of the billiard. This method allowed to greatly simplify the general expression for the Wigner current, representing its $\vp$--component as a surface integral of the product of the shifted particles wave functions. The results are also connected to an alternative approach, which takes into account the boundary conditions by adding the $\propto \delta'(x)$ term to the Hamiltonian. The latter is also generalized to the multidimensional case. 
\end{abstract}

\maketitle

\section{Introduction}
Phase space quantization \cite{polkovnikov_phase_2010, zachos_deformation_2002, curtright_concise_2014, curtright_quantum_2012, curtright_concise_2014, case_wigner_2008} is a mathematical formalism, in which the quantum systems are described not by linear operators, but scalar functions, defined on the phase space of the system --- Weyl symbols \cite{Wigner, Weyl}. The product of these functions is the so--called Moyal star product, which makes their dynamics quantum instead of classical \cite{Moyal}. The central object in the phase space quantization is the Wigner function, which is the Weyl symbol of the density matrix operator and describes the state of the quantum system.

The phase space quantization has an inherently quasiclassical nature, namely the Moyal product can be expanded in powers of $\hbar$, representing it as an ordinary multiplication with quantum corrections. In particular, in limit $\hbar \rightarrow 0$ the quantum phase space equations of motion become just classical Newtonian/Hamiltonian equations. This makes it especially convenient for studying the problem of quantum--classical correspondence \cite{dias_features_2007, berry_chaos_2001}. Analogy between the Wigner function dynamics equation and classical Liouville equation led to development of concepts of Wigner trajectories \cite{lee_wigner_1982, lee_wigner_1983, lee_wigner_1992} and Wigner current \cite{daligault_non-hamiltonian_2003, kakofengitis_wigners_2017, oliva_anharmonic_2018}. However, the $\hbar \rightarrow 0$ limit is not trivial and in general case one does not obtain the classical dynamics by its naive application \cite{berry_chaos_2001, heller_wigner_1976}.

A big open question in the field is the quantum to classical transition of systems with classically chaotic Hamiltonians. Direct application of quantum laws to them leads to the conclusion, that even a macroscopic chaotic system should become sufficiently quantum in a very short time --- tens of years or even minutes \cite{hashimoto_out--time-order_2017, cotler_out--time-order_2018}. The properties of the quantum to classical transition are also connected to the chaotic properties of the system. Namely, the non--triviality of the limit manifests itself for systems with classically chaotic dynamics \cite{berry_semi-classical_1977}. Particularly popular model systems for studying classical and quantum chaos are chaotic billiards \cite{hashimoto_out--time-order_2017, tomsovic_long-time_1993, stein_microwave_1995}, as they posses conceptually simple yet extremely rich physics. An especially fascinating phenomena in the quantum billiards are the quantum scars \cite{heller_bound-state_1984, prosen_survey_1993, berry_quantum_1989} --- excessive probability density in the phase space in the vicinity of classically unstable periodic trajectories. They are a manifestation of the connection between quantum and classical dynamical properties of physical systems.

Description of the quantum billiards within the framework of the phase space quantization requires imposing Dirichlet boundary conditions on the Wigner function. This turns out to be not particularly trivial due to non--locality of the Wigner function \cite{walton_wigner_2007, zachos_features, kryukov_infinite_2005, belchev_robin_2010, dias_wigner_2002, dias_boundaries_2021, seidov_wigner_2023}. Existence of the boundary adds an additional feature to the classical limit $\hbar \rightarrow 0$: the series expansion of the Moyal bracket in powers of $\hbar$ becomes, in general, divergent at the boundary and one has to introduce a regularization procedure \cite{liboff_quantum_2000}.

In the present paper we employ developed earlier method of imposing Dirichlet boundary conditions on the dynamics of the Wigner function \cite{seidov_wigner_2023} in order to find an analytic expression for the corresponding Wigner current. We consider the case of the particle with zero potential energy confined to some region of space --- the billiard. The results are also connected to an alternative approach \cite{dias_boundaries_2021}, which takes into account the boundary conditions by adding an extra term to the kinetic energy proportional to $\delta'(x)$ --- derivative of the Dirac delta function, which is generalized to the case of arbitrary dimensions. 

We start with the brief introduction to the phase space quantization method and the idea of the Wigner current. Next the main results are presented, i.e. the expression for the Wigner current for the particle in the billiard is derived. After that we connect our results with the proposed in \cite{dias_wigner_2002} method, which consists of including $\propto \delta'(x)$ term in the Hamiltonian. Finally the general results are applied to the case of the particle in a quantum box.

\section{Phase space quantization and the Wigner current}
Quantum mechanics in phase space is an approach in which the quantum operators are transformed to scalar functions, defined on the phase space of the system. This is done via the so--called Weyl transform, which for an operator $\hat A$ in an $n$--dimensional space ($2n$--dimensional phase space) is defined as
\begin{equation}
A(\vx, \vp) = \frac{1}{(2\pi)^n} \int e^{i \vp \vy} \left\langle \vx - \frac{\vy}{2} \right|\hat A \left| \vx + \frac{\vy}{2} \right\rangle d^n \vy.
\end{equation}
The function $A(\vx, \vp)$ is called the Weyl symbol of the operator $\hat A$. Essentially the Weyl transform is the Fourier transform of the matrix element of $\hat A$. This fact will lie in the basis of our calculations.  

The state of the quantum system is described by the Weyl symbol of the density matrix $\hat \rho$ --- the Wigner function. If the state is pure, i.e. $\hat \rho = |\varphi \rangle \langle \varphi|$, the Wigner function can be expressed via the wave function:
\begin{equation}
\begin{aligned}
W(\vx, \vp) &= \frac{1}{(2\pi)^n} \int e^{i \vp \vy} \left\langle \vx - \frac{\vy}{2} \right|\hat \rho \left| \vx + \frac{\vy}{2} \right\rangle d^n \vy =\\
&= \frac{1}{(2\pi)^n} \int e^{i \vp \vy} \Big\langle \vx - \frac{\vy}{2} \Big| \varphi \Big\rangle \Big\langle \varphi \Big| \vx + \frac{\vy}{2} \Big\rangle d^n \vy =\\
&= \frac{1}{(2\pi)^n} \int e^{i \vp \vy} \varphi^* \left(\vx - \frac{\vy}{2} \right) \varphi \left(\vx + \frac{\vy}{2} \right) d^n\vy.
\end{aligned}
\end{equation}
The Wigner function serves as a quasiprobability distribution on the phase space, in particular, its marginal distributions define the probability densities as follows:
\begin{equation}
\begin{aligned}
&\int W(\vx, \vp) d^n\vx = |\varphi(\vp)|^2 &\int W(\vx, \vp) d^n\vp = |\varphi(\vx)|^2.
\end{aligned}
\end{equation}

The product of two operators $\hat A$ and $\hat B$ is transformed to the star product of their Weyl symbols:
\begin{equation}
\hat A \hat B \rightarrow A(\vx, \vp) \star B(\vx, \vp) =  A(\vx, \vp) e^{i(\overleftarrow{\partial_{\vx}} \overrightarrow{\partial_{\vp}} - \overleftarrow{\partial_{\vp}} \overrightarrow{\partial_{\vx}})} B(\vx, \vp).
\end{equation}
The arrow above the derivative denotes the direction it acts on. Accordingly the commutator of the operators transforms to the so--called Moyal bracket:
\begin{equation}
[\hat A, \hat B] \rightarrow \{\!\{A(\vx, \vp), B(\vx, \vp) \}\!\} =A(\vx, \vp) \star B(\vx, \vp) - B(\vx, \vp) \star A(\vx, \vp).
\end{equation}

For systems with the Hamiltonian in form $H = \vp^2/2m + V(\vx)$ the dynamics of the Wigner function is then governed by the differential equation
\begin{equation}\label{eq:dotW1}
\dot W(\vx, \vp, t) = -\frac{\vp}{m} \partial_{\vx} W(\vx, \vp, t) + i \{\!\{ V(\vx), W(\vx, \vp, t) \}\!\}.
\end{equation}
This equation can be written down in integral form, which will be more suitable for our purposes:
\begin{equation}\label{eq:dotW2}
\begin{aligned}
&\dot W(\vx, \vp, t) = -\frac{\vp}{m} \partial_{\vx} W(\vx, \vp, t) + J(\vx, \vp) *_{\vp} W(\vx, \vp, t)\\
&J(\vx, \vp) = \frac{i}{(2\pi)^n} \int e^{i \vp \vy} \left\{V \left(\vx - \frac{\vy}{2}\right)  - V \left(\vx + \frac{\vy}{2}\right) \right\} d^n\vy.
\end{aligned}
\end{equation}
The $*_{\vp}$ operation denotes the convolution with respect to variable $\vp$:
\begin{equation}
f(\vp) *_p g(\vp) = \int f(\vp - \vk) g(\vk) d^n \vk.
\end{equation}
In the case of free motion, when $V(\vx) = 0$, the solution of the equation of motion is
\begin{equation}\label{eq:Wt_free}
\begin{aligned}
&W(\vx, \vp, t) = W_0\left(\vx - \frac{\vp t}{m}, \vp \right)\\
&W(\vx, \vp, 0) = W_0(\vx, \vp).
\end{aligned}
\end{equation}

The idea of the Wigner current \cite{daligault_non-hamiltonian_2003, kakofengitis_wigners_2017, oliva_anharmonic_2018} comes from the analogy with the classical probability current in the phase space. Namely, in classical case the phase space distribution function $\rho (\vx, \vp, t)$ evolves according to the Liouville equation. For the systems with Hamiltonians in form $H = \vp^2/2m + V(\vx)$ it reads
\begin{equation}
\dot{\rho} (\vx, \vp, t) = -\frac{\vp}{m} \partial_{\vx} \rho(\vx, \vp, t) - \partial_{\vx} V(\vx) \partial_{\vp} \rho (\vx, \vp, t).
\end{equation}
It can be seen as a continuity equation with the current $\vj$:
\begin{equation}
\begin{aligned}
&\dot \rho = - \nabla \cdot \vj  = - \partial_{\vx} \vj_{\vx} - \partial_{\vp} \vj_{\vp}\\
&\vj = \begin{pmatrix}
\dfrac{\vp}{m} \rho \\ -\partial_{\vx} V(\vx) \rho
\end{pmatrix}.
\end{aligned}
\end{equation}
Comparing with equations (\ref{eq:dotW1}) and (\ref{eq:dotW2}), the Wigner current is defined as
\begin{equation}
\vj_W = \begin{pmatrix}
\dfrac{\vp}{m} W \\ -i \int \{\!\{ V, W \}\!\} d\vp
\end{pmatrix} = 
\begin{pmatrix}
\dfrac{\vp}{m} W \\ -\int J *_{\vp} W d\vp
\end{pmatrix}.
\end{equation}

\section{Boundary conditions as convolution}
We consider the case of a particle confined to a billiard --- some region of space $\mathcal{B}$ in which the potential energy is zero. Formally this means, that the wave function is in form $\psi(\vx, t) = \varphi(\vx, t) B(\vx)$. Here
\begin{equation}
B(\vx) = \begin{cases}
1, &\vx \in \mathcal{B}\\
0, &\vx \notin \mathcal{B}
\end{cases}
\end{equation}
is called the indicator function of the billiard $\mathcal{B}$. The boundary condition is $\varphi(\vx \in S, t) = 0$, where $S$ is the boundary of the billiard $\mathcal{B}$ and $\varphi(\vx, t)$ is the solution of the free particle Schrodinger equation $i \dot \varphi = - \partial^2_{\vx} \varphi/(2 m)$. The wave function $\varphi(\vx, t)$ can be constructed by taking the linear combination of the eigenstates of the system in the billiard:
\begin{equation}\label{eq:phi_expansion}
\begin{aligned}
&\varphi(\vx, t) = \sum_n c_n e^{-i E_n t} \chi_n(\vx)\\
&-\frac{1}{2m} \partial^2_{\vx} \chi_n(\vx) = E_n \chi_n(\vx)\\
& \chi_n(\vx \in S) = 0.
\end{aligned}
\end{equation}  
This guaranties fulfilment of the boundary condition $\varphi(\vx \in S, t) = 0$.

\begin{figure}[h!!]
\center\includegraphics[width = 0.5\textwidth]{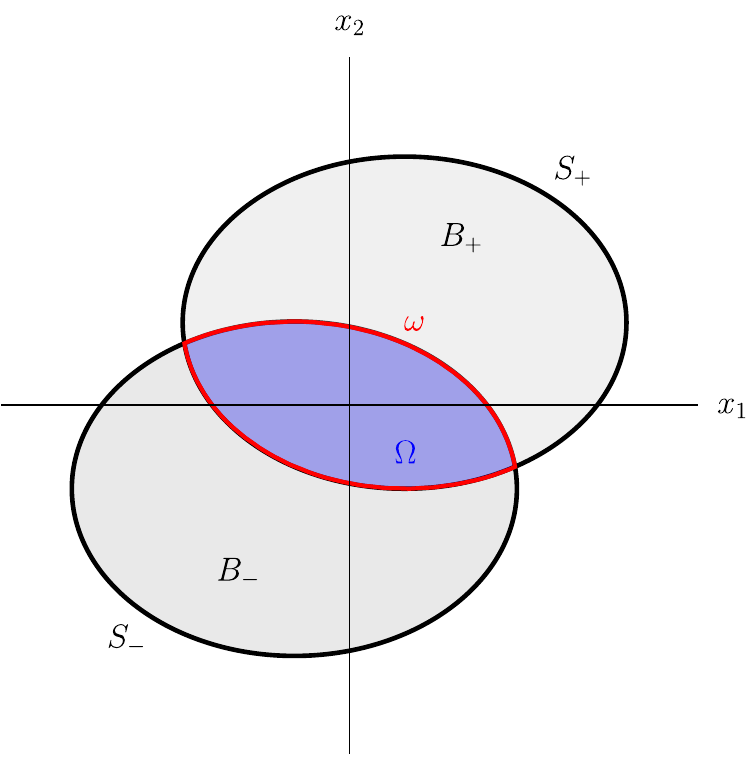}
\caption{Plot of the indicator functions of the regions and surfaces in the setup of the problem in two dimensions. Indicator functions $B_\pm$ correspond to the billiard $\mathcal{B}$ shifted by $\vy/2$ and $S_\pm$ are their surfaces. $\Omega$ is the indicator function of the intersection of the shifted billiards $\mathcal{B}_\pm$ with the surface $\omega$.}
\label{fig:billiard}
\end{figure}

By definition, the Wigner function of the particle in the billiard is
\begin{equation}
W(\vx, \vp, t) = \frac{1}{(2 \pi)^n} \int e^{i \vp \vy} \varphi^*\left(\vx - \frac{\vy}{2}, t \right) \varphi\left(\vx + \frac{\vy}{2}, t \right) B \left(\vx - \frac{\vy}{2}\right) B \left(\vx + \frac{\vy}{2}\right) d^n \vy.
\end{equation}
Next we introduce
\begin{equation}
\begin{aligned}
f(\vx, \vy, t) &= \varphi^*\left(\vx - \frac{\vy}{2}, t \right) \varphi\left(\vx + \frac{\vy}{2}, t \right)\\
\Omega(\vx, \vy) &= B \left(\vx - \frac{\vy}{2}\right) B \left(\vx + \frac{\vy}{2}\right).
\end{aligned}
\end{equation}
Here $\Omega(\vx, \vy)$ is the indicator function of the intersection of the two billiards $\mathcal{B}$, shifted by $\pm \vy/2$ and we denote its boundary as $\omega(\vx, \vy)$. The regions in two dimensions are schematically plotted in fig. \ref{fig:billiard}.

Using the convolution theorem,
\begin{equation}\label{eq:W_conv_fOmega}
W(\vx, \vp, t) = \F_y \left\{ f(\vx, \vy, t) \Omega(\vx, \vy) \right\} = \F_y \{f(\vx, \vy, t)\} *_{\vp} \F_y \{\Omega(\vx, \vy) \}.
\end{equation} 
The Fourier transform of $f(\vx, \vy, t)$ is just the Wigner function of the free particle, the dynamics of which is governed by the free solution (\ref{eq:Wt_free}). The Fourier transform of $\Omega(\vx, \vy)$ is in turn some time independent function, convolution with which imposes boundary conditions on the free particle Wigner function. This gives the Wigner function of the particle in a billiard \cite{seidov_wigner_2023}:
\begin{equation}\label{eq:W_conv}
\begin{aligned}
W(\vx, \vp, t) &= W_0 \left(\vx - \frac{\vp t}{m}, \vp \right) *_{\vp} G(\vx, \vp)\\
G(\vx, \vp) &= \F_y \{\Omega(\vx, \vy) \}.
\end{aligned}
\end{equation}

\section{Surface $\delta$--function}
In order to proceed it is required to introduce the multidimensional generalization of the $\delta$--function \cite{lange_potential_2012, lange_distribution_2015, zhang_representation_2012}. Let us consider a one dimensional interval $[a, b]$ with its indicator function
\begin{equation}
I(x) = \theta(x - a) - \theta(x - b).
\end{equation}
The derivative of the indicator function is
\begin{equation}
\partial_x I(x) = \delta(x - a) - \delta(x - b).
\end{equation}
For an arbitrary function $u(x)$ it has the property
\begin{equation}
\int u(x) \partial_x I(x) dx = \int u(x) [\delta(x - a) - \delta(x - b)] dx = u(a) - u(b),
\end{equation}
i.e. integrated with the function, the derivative of the indicator returns the sum of the functions values at the boundaries of the interval. The minus sign in front of the second term is due to the direction of the normal to the interval.  

In $n$ dimensions the boundary of the region $D$ does not consist just of two points, but is an $n-1$ dimensional surface $S$. The inward normal derivative of the indicator is called the surface delta function $\delta_S(\vx)$. Integrated with an arbitrary function, it integrates (``continuously sums'')  the values of the function at each point of the surface:
\begin{equation}
\begin{aligned}
\int u(\vx) \delta_S(\vx) d^n \vx &= \oint_S u(\vx) d^{n-1} \vx\\
\delta_S (\vx) &= \bm{n}_{\vx} \partial_{\vx} I(\vx).
\end{aligned}
\end{equation}  
Here $\bm{n}_{\vx}$ is the inward normal of $S$ at point $\vx$. One could also introduce the derivative of the surface $\delta$--function, i.e. the second derivative of the indicator:
\begin{equation}\label{eq:deltaS_prime}
\begin{aligned}
\int u(\vx) \delta'_{S}(\vx) d^n \vx &= -\oint_S \partial_{\vx} u(\vx) d^{n-1} \vx\\
\delta'_{S}(\vx) &= \partial^2_{\vx} I(\vx).
\end{aligned}
\end{equation}

\section{Equation of motion and the Wigner current}
We now aim to derive the equation of motion for the Wigner function (\ref{eq:W_conv}) and the corresponding Wigner current. To do so we calculate the derivatives with respect to time and coordinate. 

The derivative with respect to time is
\begin{equation}
\dot W(\vx, \vp, t) = -\left[ \frac{\vp}{m} \partial_{\vx} W_0 \left(\vx - \frac{\vp t}{m}, \vp \right) \right] *_{\vp} G(\vx, \vp).
\end{equation}
The convolution can be expanded using following relation:
\begin{equation} 
\begin{aligned}
[\vp u(\vp)]*_{\vp} g(\vp) &= \int (\vp-\vk) u(\vp - \vk) g(\vk) d^n \vk = \vp \int u(\vp - \vk) g(\vk) d^n \vk -\\
&- \int \vk u(\vp - \vk) g(\vk) d^n \vk = \vp [u *_{\vp} g] - u *_{\vp} [\vp g].
\end{aligned}
\end{equation}
By doing so we find
\begin{equation}
\dot W(\vx, \vp, t) = -\frac{\vp}{m} \left[\partial_{\vx} W_0\left(\vx - \frac{\vp t}{m}, \vp \right) *_{\vp} G(\vx, \vp) \right] + \frac{1}{m} \partial_{\vx} W_0\left(\vx - \frac{\vp t}{m}, \vp \right) *_{\vp} [\vp G(\vx, \vp)].
\end{equation}
The second term in the right hand side can be simplified. We use the convolution theorem ``in reverse'':
\begin{equation}
\partial_{\vx} W_0\left(\vx - \frac{\vp t}{m}, \vp \right) *_{\vp} [\vp G(\vx, \vp)] = \mathcal F_y \left\{\partial_{\vx} f(\vx, \vy, t) \mathcal F^{-1}_p \left\{\vp G(\vx, \vp) \right\}\right\}.
\end{equation}
The inverse Fourier transform is the surface delta function at the boundary $\omega(\vx, \vy)$ of the region $\Omega(\vx, \vy)$:
\begin{equation}
\begin{aligned}
\mathcal F_p^{-1} \{\vp G(\vx, \vp) \} &= i \partial_{\vy} \mathcal F^{-1}_p \{G(\vx, \vp) \} = i \partial_{\vy} \Omega(\vx, \vy) = i \delta_\omega (\vx, \vy).
\end{aligned}
\end{equation}
Thus the Fourier transform simplifies to the surface integral:
\begin{equation}
i\mathcal F_y \left\{\partial_{\vx} f(\vx, \vy, t) \delta_\omega (\vx, \vy) \right\} = \frac{i}{(2\pi)^n} \oint\limits_{\omega(\vx, \vy)} e^{i \vp \vy} \partial_{\vx} f(\vx, \vy, t) d^{n-1} \vy.
\end{equation}
The integral should be understood as follows: for a given vector $\vx$ one finds a surface $\omega(\vx, \vy)$ in the $\vy$ space and performs a usual surface integration procedure. Finally the time derivative is
\begin{equation}\label{eq:dtW}
\dot W(\vx, \vp, t) = -\frac{\vp}{m} \left[\partial_{\vx} W_0\left(\vx - \frac{\vp t}{m}, \vp \right) *_{\vp} G(\vx, \vp) \right] + \frac{i}{(2\pi)^n m} \oint\limits_{\omega(\vx, \vy)} e^{i \vp \vy} \partial_{\vx} f(\vx, \vy, t) d^{n-1} \vy.
\end{equation}

The spatial derivative is
\begin{equation}
\partial_{\vx} W(\vx, \vp, t) = \partial_{\vx} W_0 \left(\vx - \frac{\vp t}{m}, \vp \right) *_{\vp} G(\vx, \vp) + W_0 \left(\vx - \frac{\vp t}{m}, \vp \right) *_{\vp} \partial_{\vx} G(\vx, \vp).
\end{equation}
The second term turns out to be zero, which can be seen if we once again evaluate the convolution via the convolution theorem:
\begin{equation}
W_0 \left(\vx - \frac{\vp t}{m}, \vp \right) *_{\vp} \partial_{\vx} G(\vx, \vp) = \F_y \left\{f(\vx, \vy, t)\partial_{\vx} \Omega(\vx, \vy)\right\} = \F_y \left\{f(\vx, \vy, t) \delta_\omega (\vx, \vy) \right\}.
\end{equation}
This Fourier transform is zero because the function $f(\vx, \vy, t)$ is zero on $\omega(\vx, \vy)$ due to the boundary condition. Namely, $f(\vx, \vy, t) = \varphi^*(\vx - \vy/2, t)\varphi(\vx + \vy/2, t)$ and given that $\omega(\vx, \vy)$ is the boundary of the intersection of two billiards, shifted by $\pm \vy/2$, either first or second multiplier will be zero depending on the point at which the function is evaluated.

Finally, 
\begin{equation}
\partial_{\vx} W(\vx, \vp, t) = \partial_{\vx} W_0 \left(\vx - \frac{\vp t}{m}, \vp \right) *_{\vp} G(\vx, \vp)
\end{equation}
and comparing with the time derivative (\ref{eq:dtW}) we end up with the equation of motion
\begin{equation}\label{eq:W_EOM}
\dot W(\vx, \vp, t) = -\frac{\vp}{m} \partial_{\vx} W(\vx, \vp, t) + \frac{i}{(2\pi)^n m} \oint\limits_{\omega(\vx, \vy)} e^{i \vp \vy} \partial_{\vx} f(\vx, \vy, t) d^{n-1} \vy.
\end{equation}
This is a remarkable achievement --- we have drastically simplified the potential energy term in the Moyal equation (\ref{eq:dotW1}) and (\ref{eq:dotW2}). Wave functions $\varphi(\vx \pm \vy/2, t)$ in the definition of $f(\vx, \vy, t)$ are just the wave functions of the free particle, which evolve according to equation (\ref{eq:phi_expansion}). This allows a relatively simple calculation of the second term in (\ref{eq:W_EOM}) with a well defined numerical procedure. The Wigner current $\vj_W = (j_W^{\vx}\ \ j_W^{\vp})^T$ can be easily written down:
\begin{equation}
\begin{aligned}
&j_W^{\vx} = \frac{\vp}{m} W(\vx, \vp, t)\\
&\begin{aligned}
j_W^{\vp} &= -\frac{i}{(2\pi)^n m}  \int d\vp \oint\limits_{\omega(\vx, \vy)} e^{i \vp \vy} \partial_{\vx} f(\vx, \vy, t) d^{n-1} \vy =\\
&= -\frac{1}{(2\pi)^n m} \oint\limits_{\omega(\vx, \vy)} \frac{e^{i \vp \vy}}{\vy} \partial_{\vx} f(\vx, \vy, t) d^{n-1} \vy.
\end{aligned}
\end{aligned}
\end{equation}

\section{Motion in $\delta'$ potential}
The motion of a particle, reflecting from a wall at $x = 0$ in one dimension, can be described as motion with the Hamiltonian \cite{dias_wigner_2002, dias_boundaries_2021}
\begin{equation}
H = \frac{p^2}{2m} + \frac{\delta'(x)}{2m}.
\end{equation}
The additional term can be formally considered as the potential energy, but in fact it is the modification of the kinetic energy due to confinement on the positive real line \cite{dias_boundaries_2021}. We propose the generalization to the multidimensional case by replacing the one--dimensional delta--prime function with the surface delta--prime function $\delta'_S(\vx)$ at the boundary $S$ of the billiard. The multidimensional Hamiltonian then is
\begin{equation}\label{eq:H_delta}
H = \frac{\vp^2}{2m} + \frac{\delta'_S(\vx)}{2m}.
\end{equation}

In order to check that this is true, we have to demonstrate that the equation of motion (\ref{eq:dotW2}), generated by Hamiltonian (\ref{eq:H_delta}), coincides with the equation of motion (\ref{eq:W_EOM}). The first term of (\ref{eq:W_EOM}) is already in the desired form, so we have to prove that
\begin{equation}\label{eq:conv_delta_multidim}
\frac{i}{2 m} \mathcal F_y \left\{\delta'_S \left(\vx - \frac{\vy}{2} \right) - \delta'_S \left(\vx + \frac{\vy}{2} \right) \right\} *_{\vp} W(\vx, \vp, t) = \frac{i}{(2\pi)^n m} \oint\limits_{\omega(\vx, \vy)} e^{i \vp \vy} \partial_{\vx} f(\vx, \vy, t) d^{n-1} \vy.
\end{equation}
Bearing in mind, that $\F^{-1}_p \{W(\vx, \vp, t)\} = f(\vx, \vy, t) \Omega(\vx, \vp)$, see eq. (\ref{eq:W_conv_fOmega}), in the left hand side
\begin{equation}
\begin{aligned}
&\mathcal F_y \left\{\delta'_S \left(\vx - \frac{\vy}{2} \right) - \delta'_S \left(\vx + \frac{\vy}{2} \right) \right\} *_{\vp} W(\vx, \vp, t) =\\
&= \F_y \left\{\left[\delta'_S \left(\vx - \frac{\vy}{2} \right) - \delta'_S \left(\vx + \frac{\vy}{2} \right) \right] f(\vx, \vy, t) \Omega(\vx, \vy)  \right\}.
\end{aligned}
\end{equation}
Let us calculate each term separately:
\begin{equation}
\begin{aligned}
&\F_y \left\{\delta'_S \left(\vx \pm \frac{\vy}{2} \right) f(\vx, \vy, t) \Omega(\vx, \vy)  \right\} =\\
&= \frac{1}{(2 \pi)^n} \oint\limits_{S_\pm} e^{i \vp \vy} \left[\partial_{\vx} f(\vx, \vy, t) \Omega(\vx, \vy) + f(\vx, \vy, t) \delta_\omega(\vx, \vy) \right] d^{n-1} \vy.
\end{aligned}
\end{equation}
Here $S_\pm$ are the surfaces of the billiards shifted by $\pm \vy/2$, see fig. \ref{fig:billiard}. The second term in the integral is zero, because $f(\vx, \vy, t)$ is zero at the boundary of the shifted billiard. As for the first term, it is nonzero only when the indicator function of the intersection of shifted billiards $\Omega(\vx, \vy)$ is nonzero. In this case surfaces $S_\pm$ outline the intersections surface $\omega(\vx, \vy)$ (see fig. \ref{fig:billiard} for illustration), so
\begin{equation}
\oint\limits_{S_+} \Omega(\vx, \vy) + \oint\limits_{S_-} \Omega(\vx, \vy) = \oint\limits_{\omega(\vx, \vy)}.
\end{equation}
This finally leads to 
\begin{equation}
\mathcal F_y \left\{\delta'_S \left(\vx - \frac{\vy}{2} \right) - \delta'_S \left(\vx + \frac{\vy}{2} \right) \right\} *_{\vp} W(\vx, \vp, t) = \frac{1}{(2\pi)^n} \oint\limits_{\omega(\vx, \vy)} e^{i \vp \vy} \partial_{\vx} f(\vx, \vy, t) d^{n-1}\vy,
\end{equation}
which proves equality (\ref{eq:conv_delta_multidim}) and that indeed the motion in the billiard can be described by the Hamiltonian (\ref{eq:H_delta}).

\section{Particle in a box}
\begin{figure}[h!!]
\center\includegraphics[width = 0.5\textwidth]{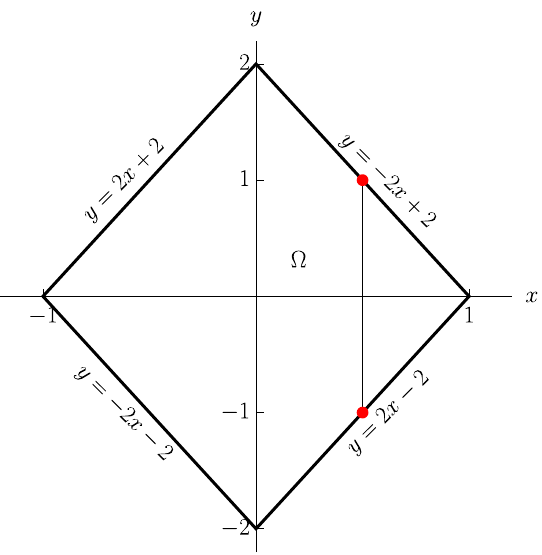}
\caption{Region $\Omega(x, y)$ of the intersection of two shifted potential boxes. Red points are the boundary of the region for the fixed value of $x$, see eq. (\ref{eq:omega_box}).}
\label{fig:Omega}
\end{figure}

In order to illustrate the general results, we consider the one--dimensional case of the quantum particle in a box. In this case the region $\Omega(x, y)$ is two dimensional and can be plotted. The corresponding contour integral with respect to variable $y$ becomes just a sum over two points. Let us set the boundaries of the box at $x = \pm 1$, then the indicator functions of the interval $[-1, 1]$ is
\begin{equation}
B(x) = \theta(x + 1) - \theta(x - 1).
\end{equation}
Next, the intersection of the shifted by $\pm y/2$ boxes is
\begin{equation}
\Omega(x, y) = B \left(x - \frac{y}{2} \right) B \left(x + \frac{y}{2} \right).
\end{equation}
Its boundary for fixed $x$ consists of two points, see fig. \ref{fig:Omega}:
\begin{equation}\label{eq:omega_box}
\omega(x, y) = \begin{cases}
\{2 x - 2, - 2 x + 2\} & 0 \leqslant x \leqslant 1\\
\{-2 x - 2, 2 x + 2\} & -1 \leqslant x < 0
\end{cases}.
\end{equation}  

\noindent Thus the contour integral over $\omega(\vx, \vy)$ reduces to a sum over the left either right pair of points depending on the sign of $x$:
\begin{equation}
\oint\limits_{\omega(\vx, \vy)} = 
\left[\theta(x) - \theta\left(x - 1 \right) \right] \sum_{y = 2 x - 2, -2 x +2} s_y + \left[\theta\left(x + 1 \right) - \theta(x) \right] \sum_{y = -2 x - 2, 2 x + 2} s_y.
\end{equation}
The value $s_y$ denotes the sign of the normal to $\Omega(x, y)$, it should be taken equal to $+1$ for the left border (lower red point in fig. \ref{fig:Omega}) and $-1$ for the right border (higher red point in fig. \ref{fig:Omega}). The $p$ component of the Wigner current becomes
\begin{equation}
\begin{aligned}\label{eq:jp_box}
j_W^p &= -\frac{1}{2 \pi m}\left[\theta(x) - \theta\left(x - 1 \right) \right] \times\\
&\times \left[\frac{e^{i p (2 x - 2)}}{2 x - 2} \partial_x f(x, y, t)\Big|_{y = 2 x - 2} - \frac{e^{i p (-2 x + 2)}}{-2 x + 2} \partial_x f(x, y, t)\Big|_{y = -2 x + 2} \right] - \\
& -\frac{1}{2 \pi m}\left[\theta\left(x + 1 \right) - \theta(x) \right] \times\\
&\times \left[\frac{e^{i p (-2 x - 2)}}{-2 x - 2} \partial_x f(x, y, t)\Big|_{y = -2 x - 2} - \frac{e^{i p (2 x + 1)}}{2 x + 2} \partial_x f(x, y, t)\Big|_{y = 2 x + 2} \right].
\end{aligned}
\end{equation}

Next we have to construct the Wigner function of the particle in a box. We start by representing the wave function as the superposition of eigenfunction of the particle in a box Hamiltonian:
\begin{equation}\label{eq:psi_box}
\begin{aligned}
&\psi(x, t) = \sum_n c_n e^{-i E_n t} \chi_n(x)\\
&\chi_n(x) = \sin \left(\frac{\pi n}{2} (x + 1) \right)\\
&E_n = \frac{\pi^2 n^2}{8}.
\end{aligned}
\end{equation}
Then the free particle Wigner function is
\begin{equation}
\begin{aligned}
&W_0(x, p, t) = \mathcal{F}_y \left\{\psi \left(x - \frac{y}{2}\right) \psi \left(x + \frac{y}{2}\right) \right\} = \sum_{n m}c_n c^*_m e^{-i (E_n - E_m) t} \lambda_{n m} (x, p)\\
&\begin{aligned}
\lambda_{nm}(x, p) &= e^{\frac{i \pi}{2}(m - n)(1 + x)} \delta(4 p + \pi (m + n)) + e^{-\frac{i \pi}{2}(m - n)(1 + x)} \delta(4 p - \pi (m + n))-\\
& - e^{\frac{i \pi}{2}(m + n)(1 + x)} \delta(4 p + \pi (m - n)) - e^{-\frac{i \pi}{2}(m + n)(1 + x)} \delta(4 p + \pi (m + n)).
\end{aligned}
\end{aligned}
\end{equation}
In order to obtain the Wigner function, the function $W_0(x, p, t)$ should be convolved with the function
\begin{equation}
\begin{aligned}
G(x, p) = \mathcal{F}_y \{ \Omega(x, y) \} &= \left[\theta(x) - \theta\left(x - 1 \right) \right] \frac{\sin(2 p (1 - x))}{\pi p} +\\
&+ \left[\theta\left(x + 1 \right) - \theta(x) \right] \frac{\sin(2 p (1 + x))}{\pi p}.
\end{aligned}
\end{equation}
Finally we obtain the Wigner function in the box
\begin{equation}\label{eq:W_box}
W(x, p, t) = W_0(x, p, t) *_p G(x, p) = \sum_{n m}c_n c^*_m e^{-i (E_n - E_m) t} \lambda_{n m} (x, p) *_p G(x, p).
\end{equation}
We do not write down the complete expression, but the convolution can be trivially calculated, given that $u(p) *_p \delta(p - a) = u(p - a)$.

\begin{figure}[h]
\subfloat[a)]{\includegraphics[width = 0.42\textwidth]{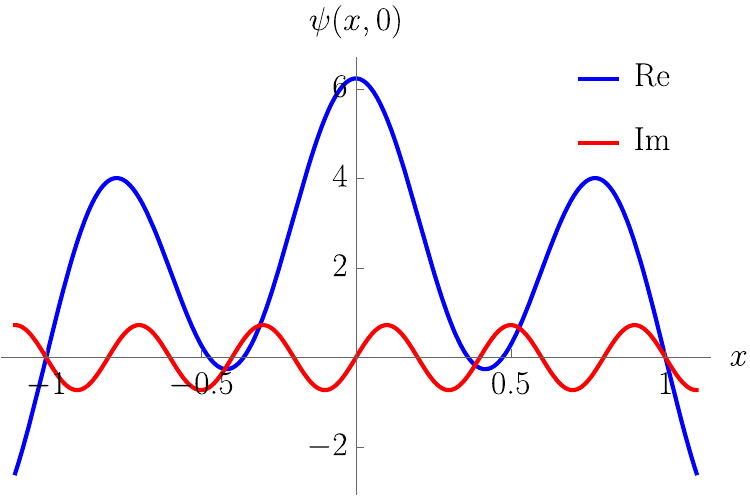}}
\subfloat[b)]{\includegraphics[width = 0.56\textwidth]{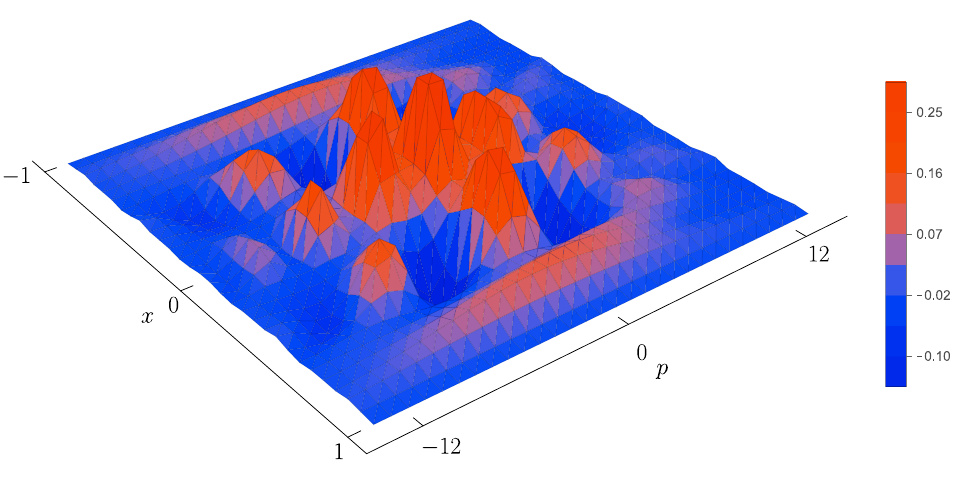}}
\caption{a) Plot of the wave function (\ref{eq:psi_xt}) with $m = 1$, $n = \{1, 5, 10\}$ and $p_0 = 5$ at $t = 0$; b) plot of the corresponding Wigner function in the box given by equation (\ref{eq:W_box}).}
\label{fig:psi}
\end{figure}

Now the Wigner current of the particle in a box can be calculated, using expression (\ref{eq:jp_box}) for $j_W^p$ and $j_W^x = p W/m$ with the Wigner function (\ref{eq:W_box}), if one makes a particular choice of coefficients $c_n$ in the wave function (\ref{eq:psi_box}). We would like to consider an initial state, localized in the center of the box with some initial momentum $p_0$. It is constructed by expanding a Gaussian in the basis of eigenfunctions $\chi_n(x)$ from (\ref{eq:psi_box}):
\begin{equation}\label{eq:psi_xt}
\begin{aligned}
&\psi(x, t) = A \sum_{\{n\}} e^{-i E_n t} \chi_n(x) \int\limits_{-1}^1 \chi_n^*(x') \phi_0(x') d x'\\
&\phi_0(x) = \left(\frac{2}{\pi}\right)^{1/4} e^{-x^2/a^2 - i p_0 x}.
\end{aligned}
\end{equation}
Here $A$ is the normalization factor and $\{n\}$ is the list with the numbers of the eigenfunctions involved in the expansion. Further we will consider an example with $m = 1$, ${n} = \{1, 5, 10\}$ and $p_0 = 5$, the corresponding wave and Wigner functions at $t = 0$ are plotted in fig. \ref{fig:psi}.

\begin{figure}[h!!]
\includegraphics[width=0.99\textwidth]{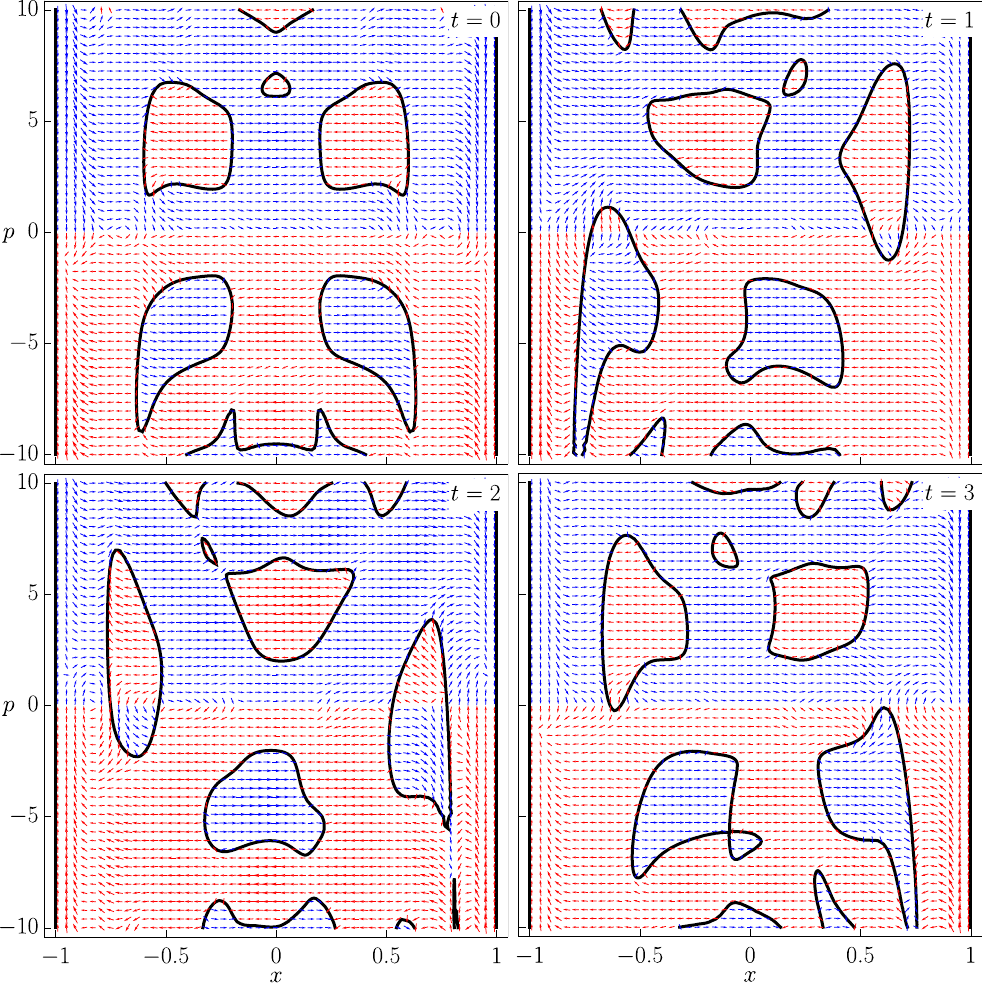}
\caption{The Wigner current in a quantum box of a particle with the wave function (\ref{eq:psi_xt}) at different times $t$. Black lines --- contours at which $W(x, p, t) = 0$. Red arrows are oriented in negative $x$ direction and blue arrows are oriented in positive $x$ direction. The Wigner current changes its direction along $x$ axis when it crosses the contours $W(x, p, t) = 0$.}
\label{fig:current}
\end{figure}

In fig. \ref{fig:current} we present plots of the Wigner current in a box for the state (\ref{eq:psi_xt}) with $n = \{1, 5, 10\}$ and $p_0 = 5$. Red arrows on the plots are facing left, i.e. for them $j_W^x < 0$ and the blue ones are facing right with $j_W^x > 0$. The black lines are the contours at which the Wigner function equals zero. One can observe, that when crossing these contours, the Wigner current changes its direction along the $x$ axis. This follows from the fact, that the Wigner function changes its sign and $j_W^x \sim W$.

\section{Conclusions}
We have derived the equation of motion for the particle in an $n$--dimensional billiard, based on the method of imposing boundary conditions to the Wigner function, defined on the entire phase space, via convolution with some function, defined by the shape of the billiard. The equation of motion allows to straightforwardly write down the Wigner current. 

The main achievement is that the expression for the $\vj_W^p$ component of the Wigner current was simplified in comparison to its general form. In particular, it is shown to be equal to a surface integral of the product $\varphi^*(\vx - \vy/2, t) \varphi(\vx + \vy/2, t)$ of the particles wave functions on the boundary of the region $\Omega(\vx, \vy)$ --- the intersection of shifted by $\pm \vy/2$ copies of the billiard. In one dimensional case it is especially simple, because the surface integral becomes just a discrete sum with several terms.

Additionally, it is shown, that the convolution method of imposing boundary conditions is equivalent to an alternative approach in which one adds a term $\propto \delta'(x)$ to the kinetic energy. This method, previously proposed only in a single dimension, is generalized to the multidimensional case. 

\section{Acknowledgments}
The publication was prepared within the framework of the Academic Fund Program at HSE University \#24-00-038 ``Quasiclassical dynamics of quantum systems: chaotic and spatially nonhomogeneous like heterostructures superconductor--magnetic''.


\begin{thebibliography}{36}%
\makeatletter
\providecommand \@ifxundefined [1]{%
 \@ifx{#1\undefined}
}%
\providecommand \@ifnum [1]{%
 \ifnum #1\expandafter \@firstoftwo
 \else \expandafter \@secondoftwo
 \fi
}%
\providecommand \@ifx [1]{%
 \ifx #1\expandafter \@firstoftwo
 \else \expandafter \@secondoftwo
 \fi
}%
\providecommand \natexlab [1]{#1}%
\providecommand \enquote  [1]{``#1''}%
\providecommand \bibnamefont  [1]{#1}%
\providecommand \bibfnamefont [1]{#1}%
\providecommand \citenamefont [1]{#1}%
\providecommand \href@noop [0]{\@secondoftwo}%
\providecommand \href [0]{\begingroup \@sanitize@url \@href}%
\providecommand \@href[1]{\@@startlink{#1}\@@href}%
\providecommand \@@href[1]{\endgroup#1\@@endlink}%
\providecommand \@sanitize@url [0]{\catcode `\\12\catcode `\$12\catcode
  `\&12\catcode `\#12\catcode `\^12\catcode `\_12\catcode `\%12\relax}%
\providecommand \@@startlink[1]{}%
\providecommand \@@endlink[0]{}%
\providecommand \url  [0]{\begingroup\@sanitize@url \@url }%
\providecommand \@url [1]{\endgroup\@href {#1}{\urlprefix }}%
\providecommand \urlprefix  [0]{URL }%
\providecommand \Eprint [0]{\href }%
\providecommand \doibase [0]{https://doi.org/}%
\providecommand \selectlanguage [0]{\@gobble}%
\providecommand \bibinfo  [0]{\@secondoftwo}%
\providecommand \bibfield  [0]{\@secondoftwo}%
\providecommand \translation [1]{[#1]}%
\providecommand \BibitemOpen [0]{}%
\providecommand \bibitemStop [0]{}%
\providecommand \bibitemNoStop [0]{.\EOS\space}%
\providecommand \EOS [0]{\spacefactor3000\relax}%
\providecommand \BibitemShut  [1]{\csname bibitem#1\endcsname}%
\let\auto@bib@innerbib\@empty
\bibitem [{\citenamefont {Polkovnikov}(2010)}]{polkovnikov_phase_2010}%
  \BibitemOpen
  \bibfield  {author} {\bibinfo {author} {\bibfnamefont {A.}~\bibnamefont
  {Polkovnikov}},\ }\bibfield  {title} {\bibinfo {title} {Phase space
  representation of quantum dynamics},\ }\href
  {https://doi.org/10.1016/j.aop.2010.02.006} {\bibfield  {journal} {\bibinfo
  {journal} {Annals of Physics}\ }\textbf {\bibinfo {volume} {325}},\ \bibinfo
  {pages} {1790} (\bibinfo {year} {2010})}\BibitemShut {NoStop}%
\bibitem [{\citenamefont {Zachos}(2002)}]{zachos_deformation_2002}%
  \BibitemOpen
  \bibfield  {author} {\bibinfo {author} {\bibfnamefont {C.}~\bibnamefont
  {Zachos}},\ }\bibfield  {title} {\bibinfo {title} {Deformation quantization:
  Quantum mechanics lives and works in phase-space},\ }\href
  {https://doi.org/10.1142/S0217751X02006079} {\bibfield  {journal} {\bibinfo
  {journal} {International Journal of Modern Physics A}\ }\textbf {\bibinfo
  {volume} {17}},\ \bibinfo {pages} {297} (\bibinfo {year} {2002})}\BibitemShut
  {NoStop}%
\bibitem [{\citenamefont {Curtright}\ \emph {et~al.}(2014)\citenamefont
  {Curtright}, \citenamefont {Fairlie},\ and\ \citenamefont
  {Zachos}}]{curtright_concise_2014}%
  \BibitemOpen
  \bibfield  {author} {\bibinfo {author} {\bibfnamefont {T.}~\bibnamefont
  {Curtright}}, \bibinfo {author} {\bibfnamefont {D.}~\bibnamefont {Fairlie}},\
  and\ \bibinfo {author} {\bibfnamefont {C.}~\bibnamefont {Zachos}},\
  }\href@noop {} {\emph {\bibinfo {title} {A concise treatise on quantum
  mechanics in phase space}}}\ (\bibinfo  {publisher} {World Scientific},\
  \bibinfo {address} {New Jersey},\ \bibinfo {year} {2014})\BibitemShut
  {NoStop}%
\bibitem [{\citenamefont {Curtright}\ and\ \citenamefont
  {Zachos}(2012)}]{curtright_quantum_2012}%
  \BibitemOpen
  \bibfield  {author} {\bibinfo {author} {\bibfnamefont {T.~L.}\ \bibnamefont
  {Curtright}}\ and\ \bibinfo {author} {\bibfnamefont {C.~K.}\ \bibnamefont
  {Zachos}},\ }\bibfield  {title} {\bibinfo {title} {Quantum {Mechanics} in
  {Phase} {Space}},\ }\href {https://doi.org/10.1142/S2251158X12000069}
  {\bibfield  {journal} {\bibinfo  {journal} {Asia Pacific Physics Newsletter}\
  }\textbf {\bibinfo {volume} {01}},\ \bibinfo {pages} {37} (\bibinfo {year}
  {2012})}\BibitemShut {NoStop}%
\bibitem [{\citenamefont {Case}(2008)}]{case_wigner_2008}%
  \BibitemOpen
  \bibfield  {author} {\bibinfo {author} {\bibfnamefont {W.~B.}\ \bibnamefont
  {Case}},\ }\bibfield  {title} {\bibinfo {title} {Wigner functions and {Weyl}
  transforms for pedestrians},\ }\href {https://doi.org/10.1119/1.2957889}
  {\bibfield  {journal} {\bibinfo  {journal} {American Journal of Physics}\
  }\textbf {\bibinfo {volume} {76}},\ \bibinfo {pages} {937} (\bibinfo {year}
  {2008})}\BibitemShut {NoStop}%
\bibitem [{\citenamefont {Wigner}(1932)}]{Wigner}%
  \BibitemOpen
  \bibfield  {author} {\bibinfo {author} {\bibfnamefont {E.}~\bibnamefont
  {Wigner}},\ }\bibfield  {title} {\bibinfo {title} {On the {Quantum}
  {Correction} {For} {Thermodynamic} {Equilibrium}},\ }\href
  {https://doi.org/10.1103/PhysRev.40.749} {\bibfield  {journal} {\bibinfo
  {journal} {Physical Review}\ }\textbf {\bibinfo {volume} {40}},\ \bibinfo
  {pages} {749} (\bibinfo {year} {1932})}\BibitemShut {NoStop}%
\bibitem [{\citenamefont {Weyl}(1927)}]{Weyl}%
  \BibitemOpen
  \bibfield  {author} {\bibinfo {author} {\bibfnamefont {H.}~\bibnamefont
  {Weyl}},\ }\bibfield  {title} {\bibinfo {title} {Quantenmechanik und
  {Gruppentheorie}},\ }\href {https://doi.org/10.1007/BF02055756} {\bibfield
  {journal} {\bibinfo  {journal} {Zeitschrift für Physik}\ }\textbf {\bibinfo
  {volume} {46}},\ \bibinfo {pages} {1} (\bibinfo {year} {1927})}\BibitemShut
  {NoStop}%
\bibitem [{\citenamefont {Moyal}(1949)}]{Moyal}%
  \BibitemOpen
  \bibfield  {author} {\bibinfo {author} {\bibfnamefont {J.~E.}\ \bibnamefont
  {Moyal}},\ }\bibfield  {title} {\bibinfo {title} {Quantum mechanics as a
  statistical theory},\ }\href {https://doi.org/10.1017/S0305004100000487}
  {\bibfield  {journal} {\bibinfo  {journal} {Mathematical Proceedings of the
  Cambridge Philosophical Society}\ }\textbf {\bibinfo {volume} {45}},\
  \bibinfo {pages} {99} (\bibinfo {year} {1949})}\BibitemShut {NoStop}%
\bibitem [{\citenamefont {Dias}\ and\ \citenamefont
  {Prata}(2007)}]{dias_features_2007}%
  \BibitemOpen
  \bibfield  {author} {\bibinfo {author} {\bibfnamefont {N.~C.}\ \bibnamefont
  {Dias}}\ and\ \bibinfo {author} {\bibfnamefont {J.~N.}\ \bibnamefont
  {Prata}},\ }\bibfield  {title} {\bibinfo {title} {Features of {Moyal}
  trajectories},\ }\href {https://doi.org/10.1063/1.2409495} {\bibfield
  {journal} {\bibinfo  {journal} {Journal of Mathematical Physics}\ }\textbf
  {\bibinfo {volume} {48}},\ \bibinfo {pages} {012109} (\bibinfo {year}
  {2007})}\BibitemShut {NoStop}%
\bibitem [{\citenamefont {Berry}(2001)}]{berry_chaos_2001}%
  \BibitemOpen
  \bibfield  {author} {\bibinfo {author} {\bibfnamefont {M.}~\bibnamefont
  {Berry}},\ }\bibfield  {title} {\bibinfo {title} {Chaos and the semiclassical
  limit of quantum mechanics (is the moon there when somebody looks?)},\
  }\href@noop {} {\bibfield  {journal} {\bibinfo  {journal} {Quantum Mechanics:
  Scientific perspectives on divine action}\ ,\ \bibinfo {pages} {41}}
  (\bibinfo {year} {2001})}\BibitemShut {NoStop}%
\bibitem [{\citenamefont {Lee}\ and\ \citenamefont
  {Scully}(1982)}]{lee_wigner_1982}%
  \BibitemOpen
  \bibfield  {author} {\bibinfo {author} {\bibfnamefont {H.-W.}\ \bibnamefont
  {Lee}}\ and\ \bibinfo {author} {\bibfnamefont {M.~O.}\ \bibnamefont
  {Scully}},\ }\bibfield  {title} {\bibinfo {title} {Wigner phase-space
  description of a {Morse} oscillator},\ }\href
  {https://doi.org/10.1063/1.444412} {\bibfield  {journal} {\bibinfo  {journal}
  {The Journal of Chemical Physics}\ }\textbf {\bibinfo {volume} {77}},\
  \bibinfo {pages} {4604} (\bibinfo {year} {1982})}\BibitemShut {NoStop}%
\bibitem [{\citenamefont {Lee}\ and\ \citenamefont
  {Scully}(1983)}]{lee_wigner_1983}%
  \BibitemOpen
  \bibfield  {author} {\bibinfo {author} {\bibfnamefont {H.-W.}\ \bibnamefont
  {Lee}}\ and\ \bibinfo {author} {\bibfnamefont {M.~O.}\ \bibnamefont
  {Scully}},\ }\bibfield  {title} {\bibinfo {title} {The {Wigner} phase-space
  description of collision processes},\ }\href
  {https://doi.org/10.1007/BF01889411} {\bibfield  {journal} {\bibinfo
  {journal} {Foundations of Physics}\ }\textbf {\bibinfo {volume} {13}},\
  \bibinfo {pages} {61} (\bibinfo {year} {1983})}\BibitemShut {NoStop}%
\bibitem [{\citenamefont {Lee}(1992)}]{lee_wigner_1992}%
  \BibitemOpen
  \bibfield  {author} {\bibinfo {author} {\bibfnamefont {H.-W.}\ \bibnamefont
  {Lee}},\ }\bibfield  {title} {\bibinfo {title} {Wigner trajectories of a
  {Gaussian} wave packet perturbed by a weak potential},\ }\href
  {https://doi.org/10.1007/BF00733392} {\bibfield  {journal} {\bibinfo
  {journal} {Foundations of Physics}\ }\textbf {\bibinfo {volume} {22}},\
  \bibinfo {pages} {995} (\bibinfo {year} {1992})}\BibitemShut {NoStop}%
\bibitem [{\citenamefont {Daligault}(2003)}]{daligault_non-hamiltonian_2003}%
  \BibitemOpen
  \bibfield  {author} {\bibinfo {author} {\bibfnamefont {J.}~\bibnamefont
  {Daligault}},\ }\bibfield  {title} {\bibinfo {title} {Non-{Hamiltonian}
  dynamics and trajectory methods in quantum phase spaces},\ }\href
  {https://doi.org/10.1103/PhysRevA.68.010501} {\bibfield  {journal} {\bibinfo
  {journal} {Physical Review A}\ }\textbf {\bibinfo {volume} {68}},\ \bibinfo
  {pages} {010501} (\bibinfo {year} {2003})}\BibitemShut {NoStop}%
\bibitem [{\citenamefont {Kakofengitis}\ and\ \citenamefont
  {Steuernagel}(2017)}]{kakofengitis_wigners_2017}%
  \BibitemOpen
  \bibfield  {author} {\bibinfo {author} {\bibfnamefont {D.}~\bibnamefont
  {Kakofengitis}}\ and\ \bibinfo {author} {\bibfnamefont {O.}~\bibnamefont
  {Steuernagel}},\ }\bibfield  {title} {\bibinfo {title} {Wigner’s quantum
  phase-space current in weakly-anharmonic weakly-excited two-state systems},\
  }\href {https://doi.org/10.1140/epjp/i2017-11634-2} {\bibfield  {journal}
  {\bibinfo  {journal} {The European Physical Journal Plus}\ }\textbf {\bibinfo
  {volume} {132}},\ \bibinfo {pages} {381} (\bibinfo {year}
  {2017})}\BibitemShut {NoStop}%
\bibitem [{\citenamefont {Oliva}\ \emph {et~al.}(2018)\citenamefont {Oliva},
  \citenamefont {Kakofengitis},\ and\ \citenamefont
  {Steuernagel}}]{oliva_anharmonic_2018}%
  \BibitemOpen
  \bibfield  {author} {\bibinfo {author} {\bibfnamefont {M.}~\bibnamefont
  {Oliva}}, \bibinfo {author} {\bibfnamefont {D.}~\bibnamefont
  {Kakofengitis}},\ and\ \bibinfo {author} {\bibfnamefont {O.}~\bibnamefont
  {Steuernagel}},\ }\bibfield  {title} {\bibinfo {title} {Anharmonic quantum
  mechanical systems do not feature phase space trajectories},\ }\href
  {https://doi.org/10.1016/j.physa.2017.10.047} {\bibfield  {journal} {\bibinfo
   {journal} {Physica A: Statistical Mechanics and its Applications}\ }\textbf
  {\bibinfo {volume} {502}},\ \bibinfo {pages} {201} (\bibinfo {year}
  {2018})}\BibitemShut {NoStop}%
\bibitem [{\citenamefont {Heller}(1976)}]{heller_wigner_1976}%
  \BibitemOpen
  \bibfield  {author} {\bibinfo {author} {\bibfnamefont {E.~J.}\ \bibnamefont
  {Heller}},\ }\bibfield  {title} {\bibinfo {title} {Wigner phase space method:
  {Analysis} for semiclassical applications},\ }\href
  {https://doi.org/10.1063/1.433238} {\bibfield  {journal} {\bibinfo  {journal}
  {The Journal of Chemical Physics}\ }\textbf {\bibinfo {volume} {65}},\
  \bibinfo {pages} {1289} (\bibinfo {year} {1976})}\BibitemShut {NoStop}%
\bibitem [{\citenamefont {Hashimoto}\ \emph {et~al.}(2017)\citenamefont
  {Hashimoto}, \citenamefont {Murata},\ and\ \citenamefont
  {Yoshii}}]{hashimoto_out--time-order_2017}%
  \BibitemOpen
  \bibfield  {author} {\bibinfo {author} {\bibfnamefont {K.}~\bibnamefont
  {Hashimoto}}, \bibinfo {author} {\bibfnamefont {K.}~\bibnamefont {Murata}},\
  and\ \bibinfo {author} {\bibfnamefont {R.}~\bibnamefont {Yoshii}},\
  }\bibfield  {title} {\bibinfo {title} {Out-of-time-order correlators in
  quantum mechanics},\ }\href {https://doi.org/10.1007/JHEP10(2017)138}
  {\bibfield  {journal} {\bibinfo  {journal} {Journal of High Energy Physics}\
  }\textbf {\bibinfo {volume} {2017}},\ \bibinfo {pages} {138} (\bibinfo {year}
  {2017})}\BibitemShut {NoStop}%
\bibitem [{\citenamefont {Cotler}\ \emph {et~al.}(2018)\citenamefont {Cotler},
  \citenamefont {Ding},\ and\ \citenamefont
  {Penington}}]{cotler_out--time-order_2018}%
  \BibitemOpen
  \bibfield  {author} {\bibinfo {author} {\bibfnamefont {J.~S.}\ \bibnamefont
  {Cotler}}, \bibinfo {author} {\bibfnamefont {D.}~\bibnamefont {Ding}},\ and\
  \bibinfo {author} {\bibfnamefont {G.~R.}\ \bibnamefont {Penington}},\
  }\bibfield  {title} {\bibinfo {title} {Out-of-time-order operators and the
  butterfly effect},\ }\href {https://doi.org/10.1016/j.aop.2018.07.020}
  {\bibfield  {journal} {\bibinfo  {journal} {Annals of Physics}\ }\textbf
  {\bibinfo {volume} {396}},\ \bibinfo {pages} {318} (\bibinfo {year}
  {2018})}\BibitemShut {NoStop}%
\bibitem [{\citenamefont {Berry}(1977)}]{berry_semi-classical_1977}%
  \BibitemOpen
  \bibfield  {author} {\bibinfo {author} {\bibfnamefont {M.~V.}\ \bibnamefont
  {Berry}},\ }\bibfield  {title} {\bibinfo {title} {Semi-{Classical}
  {Mechanics} in {Phase} {Space}: {A} {Study} of {Wigner}'s {Function}},\
  }\href {https://doi.org/10.1098/rsta.1977.0145} {\bibfield  {journal}
  {\bibinfo  {journal} {Philosophical Transactions of the Royal Society of
  London Series A}\ }\textbf {\bibinfo {volume} {287}},\ \bibinfo {pages} {237}
  (\bibinfo {year} {1977})},\ \bibinfo {note} {aDS Bibcode:
  1977RSPTA.287..237B}\BibitemShut {NoStop}%
\bibitem [{\citenamefont {Tomsovic}\ and\ \citenamefont
  {Heller}(1993)}]{tomsovic_long-time_1993}%
  \BibitemOpen
  \bibfield  {author} {\bibinfo {author} {\bibfnamefont {S.}~\bibnamefont
  {Tomsovic}}\ and\ \bibinfo {author} {\bibfnamefont {E.~J.}\ \bibnamefont
  {Heller}},\ }\bibfield  {title} {\bibinfo {title} {Long-time semiclassical
  dynamics of chaos: {The} stadium billiard},\ }\href
  {https://doi.org/10.1103/PhysRevE.47.282} {\bibfield  {journal} {\bibinfo
  {journal} {Physical Review E}\ }\textbf {\bibinfo {volume} {47}},\ \bibinfo
  {pages} {282} (\bibinfo {year} {1993})}\BibitemShut {NoStop}%
\bibitem [{\citenamefont {Stein}\ \emph {et~al.}(1995)\citenamefont {Stein},
  \citenamefont {Stöckmann},\ and\ \citenamefont
  {Stoffregen}}]{stein_microwave_1995}%
  \BibitemOpen
  \bibfield  {author} {\bibinfo {author} {\bibfnamefont {J.}~\bibnamefont
  {Stein}}, \bibinfo {author} {\bibfnamefont {H.-J.}\ \bibnamefont
  {Stöckmann}},\ and\ \bibinfo {author} {\bibfnamefont {U.}~\bibnamefont
  {Stoffregen}},\ }\bibfield  {title} {\bibinfo {title} {Microwave {Studies} of
  {Billiard} {Green} {Functions} and {Propagators}},\ }\href
  {https://doi.org/10.1103/PhysRevLett.75.53} {\bibfield  {journal} {\bibinfo
  {journal} {Physical Review Letters}\ }\textbf {\bibinfo {volume} {75}},\
  \bibinfo {pages} {53} (\bibinfo {year} {1995})}\BibitemShut {NoStop}%
\bibitem [{\citenamefont {Heller}(1984)}]{heller_bound-state_1984}%
  \BibitemOpen
  \bibfield  {author} {\bibinfo {author} {\bibfnamefont {E.~J.}\ \bibnamefont
  {Heller}},\ }\bibfield  {title} {\bibinfo {title} {Bound-{State}
  {Eigenfunctions} of {Classically} {Chaotic} {Hamiltonian} {Systems}: {Scars}
  of {Periodic} {Orbits}},\ }\href
  {https://doi.org/10.1103/PhysRevLett.53.1515} {\bibfield  {journal} {\bibinfo
   {journal} {Physical Review Letters}\ }\textbf {\bibinfo {volume} {53}},\
  \bibinfo {pages} {1515} (\bibinfo {year} {1984})}\BibitemShut {NoStop}%
\bibitem [{\citenamefont {Prosen}\ and\ \citenamefont
  {Robnik}(1993)}]{prosen_survey_1993}%
  \BibitemOpen
  \bibfield  {author} {\bibinfo {author} {\bibfnamefont {T.}~\bibnamefont
  {Prosen}}\ and\ \bibinfo {author} {\bibfnamefont {M.}~\bibnamefont
  {Robnik}},\ }\bibfield  {title} {\bibinfo {title} {Survey of the
  eigenfunctions of a billiard system between integrability and chaos},\ }\href
  {https://doi.org/10.1088/0305-4470/26/20/021} {\bibfield  {journal} {\bibinfo
   {journal} {Journal of Physics A: Mathematical and General}\ }\textbf
  {\bibinfo {volume} {26}},\ \bibinfo {pages} {5365} (\bibinfo {year}
  {1993})}\BibitemShut {NoStop}%
\bibitem [{\citenamefont {Berry}(1989)}]{berry_quantum_1989}%
  \BibitemOpen
  \bibfield  {author} {\bibinfo {author} {\bibfnamefont {M.~V.}\ \bibnamefont
  {Berry}},\ }\bibfield  {title} {\bibinfo {title} {Quantum {Scars} of
  {Classical} {Closed} {Orbits} in {Phase} {Space}},\ }\href
  {https://doi.org/10.1098/rspa.1989.0052} {\bibfield  {journal} {\bibinfo
  {journal} {Proceedings of the Royal Society of London Series A}\ }\textbf
  {\bibinfo {volume} {423}},\ \bibinfo {pages} {219} (\bibinfo {year}
  {1989})},\ \bibinfo {note} {aDS Bibcode: 1989RSPSA.423..219B}\BibitemShut
  {NoStop}%
\bibitem [{\citenamefont {Walton}(2007)}]{walton_wigner_2007}%
  \BibitemOpen
  \bibfield  {author} {\bibinfo {author} {\bibfnamefont {M.~A.}\ \bibnamefont
  {Walton}},\ }\bibfield  {title} {\bibinfo {title} {Wigner functions, contact
  interactions, and matching},\ }\href
  {https://doi.org/10.1016/j.aop.2006.11.015} {\bibfield  {journal} {\bibinfo
  {journal} {Annals of Physics}\ }\textbf {\bibinfo {volume} {322}},\ \bibinfo
  {pages} {2233} (\bibinfo {year} {2007})}\BibitemShut {NoStop}%
\bibitem [{\citenamefont {Curtright}\ \emph {et~al.}(1998)\citenamefont
  {Curtright}, \citenamefont {Fairlie},\ and\ \citenamefont
  {Zachos}}]{zachos_features}%
  \BibitemOpen
  \bibfield  {author} {\bibinfo {author} {\bibfnamefont {T.}~\bibnamefont
  {Curtright}}, \bibinfo {author} {\bibfnamefont {D.}~\bibnamefont {Fairlie}},\
  and\ \bibinfo {author} {\bibfnamefont {C.}~\bibnamefont {Zachos}},\
  }\bibfield  {title} {\bibinfo {title} {Features of time-independent wigner
  functions},\ }\href {https://doi.org/10.1103/PhysRevD.58.025002} {\bibfield
  {journal} {\bibinfo  {journal} {Phys. Rev. D}\ }\textbf {\bibinfo {volume}
  {58}},\ \bibinfo {pages} {025002} (\bibinfo {year} {1998})}\BibitemShut
  {NoStop}%
\bibitem [{\citenamefont {Kryukov}\ and\ \citenamefont
  {Walton}(2005)}]{kryukov_infinite_2005}%
  \BibitemOpen
  \bibfield  {author} {\bibinfo {author} {\bibfnamefont {S.}~\bibnamefont
  {Kryukov}}\ and\ \bibinfo {author} {\bibfnamefont {M.}~\bibnamefont
  {Walton}},\ }\bibfield  {title} {\bibinfo {title} {On infinite walls in
  deformation quantization},\ }\href
  {https://doi.org/10.1016/j.aop.2004.12.004} {\bibfield  {journal} {\bibinfo
  {journal} {Annals of Physics}\ }\textbf {\bibinfo {volume} {317}},\ \bibinfo
  {pages} {474} (\bibinfo {year} {2005})}\BibitemShut {NoStop}%
\bibitem [{\citenamefont {Belchev}\ and\ \citenamefont
  {Walton}(2010)}]{belchev_robin_2010}%
  \BibitemOpen
  \bibfield  {author} {\bibinfo {author} {\bibfnamefont {B.}~\bibnamefont
  {Belchev}}\ and\ \bibinfo {author} {\bibfnamefont {M.~A.}\ \bibnamefont
  {Walton}},\ }\bibfield  {title} {\bibinfo {title} {On {Robin} boundary
  conditions and the {Morse} potential in quantum mechanics},\ }\href
  {https://doi.org/10.1088/1751-8113/43/8/085301} {\bibfield  {journal}
  {\bibinfo  {journal} {Journal of Physics A: Mathematical and Theoretical}\
  }\textbf {\bibinfo {volume} {43}},\ \bibinfo {pages} {085301} (\bibinfo
  {year} {2010})}\BibitemShut {NoStop}%
\bibitem [{\citenamefont {Dias}\ and\ \citenamefont
  {Prata}(2002)}]{dias_wigner_2002}%
  \BibitemOpen
  \bibfield  {author} {\bibinfo {author} {\bibfnamefont {N.~C.}\ \bibnamefont
  {Dias}}\ and\ \bibinfo {author} {\bibfnamefont {J.~N.}\ \bibnamefont
  {Prata}},\ }\bibfield  {title} {\bibinfo {title} {Wigner functions with
  boundaries},\ }\href {https://doi.org/10.1063/1.1504885} {\bibfield
  {journal} {\bibinfo  {journal} {Journal of Mathematical Physics}\ }\textbf
  {\bibinfo {volume} {43}},\ \bibinfo {pages} {4602} (\bibinfo {year}
  {2002})}\BibitemShut {NoStop}%
\bibitem [{\citenamefont {Dias}\ and\ \citenamefont
  {Prata}(2021)}]{dias_boundaries_2021}%
  \BibitemOpen
  \bibfield  {author} {\bibinfo {author} {\bibfnamefont {N.~C.}\ \bibnamefont
  {Dias}}\ and\ \bibinfo {author} {\bibfnamefont {J.~N.}\ \bibnamefont
  {Prata}},\ }\bibfield  {title} {\bibinfo {title} {Boundaries and profiles in
  the {Wigner} formalism},\ }\href {https://doi.org/10.1007/s10825-021-01803-7}
  {\bibfield  {journal} {\bibinfo  {journal} {Journal of Computational
  Electronics}\ }\textbf {\bibinfo {volume} {20}},\ \bibinfo {pages} {2020}
  (\bibinfo {year} {2021})}\BibitemShut {NoStop}%
\bibitem [{\citenamefont {Seidov}(2023)}]{seidov_wigner_2023}%
  \BibitemOpen
  \bibfield  {author} {\bibinfo {author} {\bibfnamefont {S.~S.}\ \bibnamefont
  {Seidov}},\ }\bibfield  {title} {\bibinfo {title} {Wigner function dynamics
  with boundaries expressed as convolution},\ }\href
  {https://doi.org/10.1088/1751-8121/ace6e5} {\bibfield  {journal} {\bibinfo
  {journal} {Journal of Physics A: Mathematical and Theoretical}\ }\textbf
  {\bibinfo {volume} {56}},\ \bibinfo {pages} {325303} (\bibinfo {year}
  {2023})}\BibitemShut {NoStop}%
\bibitem [{\citenamefont {Liboff}(2000)}]{liboff_quantum_2000}%
  \BibitemOpen
  \bibfield  {author} {\bibinfo {author} {\bibfnamefont {R.~L.}\ \bibnamefont
  {Liboff}},\ }\bibfield  {title} {\bibinfo {title} {Quantum billiard chaos},\
  }\href {https://doi.org/10.1016/S0375-9601(00)00224-3} {\bibfield  {journal}
  {\bibinfo  {journal} {Physics Letters A}\ }\textbf {\bibinfo {volume}
  {269}},\ \bibinfo {pages} {230} (\bibinfo {year} {2000})}\BibitemShut
  {NoStop}%
\bibitem [{\citenamefont {Lange}(2012)}]{lange_potential_2012}%
  \BibitemOpen
  \bibfield  {author} {\bibinfo {author} {\bibfnamefont {R.-J.}\ \bibnamefont
  {Lange}},\ }\bibfield  {title} {\bibinfo {title} {Potential theory, path
  integrals and the {Laplacian} of the indicator},\ }\href
  {https://doi.org/10.1007/JHEP11(2012)032} {\bibfield  {journal} {\bibinfo
  {journal} {Journal of High Energy Physics}\ }\textbf {\bibinfo {volume}
  {2012}},\ \bibinfo {pages} {32} (\bibinfo {year} {2012})}\BibitemShut
  {NoStop}%
\bibitem [{\citenamefont {Lange}(2015)}]{lange_distribution_2015}%
  \BibitemOpen
  \bibfield  {author} {\bibinfo {author} {\bibfnamefont {R.-J.}\ \bibnamefont
  {Lange}},\ }\bibfield  {title} {\bibinfo {title} {Distribution theory for
  {Schrödinger}’s integral equation},\ }\href
  {https://doi.org/10.1063/1.4936302} {\bibfield  {journal} {\bibinfo
  {journal} {Journal of Mathematical Physics}\ }\textbf {\bibinfo {volume}
  {56}},\ \bibinfo {pages} {122105} (\bibinfo {year} {2015})}\BibitemShut
  {NoStop}%
\bibitem [{\citenamefont {Zhang}\ and\ \citenamefont
  {Zheng}(2012)}]{zhang_representation_2012}%
  \BibitemOpen
  \bibfield  {author} {\bibinfo {author} {\bibfnamefont {Z.}~\bibnamefont
  {Zhang}}\ and\ \bibinfo {author} {\bibfnamefont {X.}~\bibnamefont {Zheng}},\
  }\bibfield  {title} {\bibinfo {title} {The {Representation} of {Line} {Dirac}
  {Delta} {Function} {Along} a {Space} {Curve}}\ }\href
  {https://doi.org/10.48550/arxiv.1209.3221} {10.48550/arxiv.1209.3221}
  (\bibinfo {year} {2012})\BibitemShut {NoStop}%
\end{thebibliography}
%

\end{document}